\documentclass[11pt,twoside]{article}

\usepackage{graphicx}
\usepackage{asp2006}
\usepackage{epsf}
\usepackage{lscape}

\begin{document}
\title{Properties of 513 isolated galaxies in the Local Supercluster}
\author{I. D. Karachentsev and D. I. Makarov}
\affil{Special Astrophysical Observatory  Russian Academy Sciences,
Nizhnij Arkhyz, Russia}
\author{V. E. Karachentseva}
\affil{Main Astronomical Observatory,
National Academy of Sciences of Ukraine}
\author{O. V. Melnyk}
\affil{Astronomical Observatory, Kiev National University, Ukraine}

\begin{abstract}
We introduce the first entire-sky catalog of the most isolated
nearby galaxies with radial velocities $V_{LG} < 3500$ km/s.
This kind of
cosmic ``orphans'' amounts to 4\% among all known galaxies within the same
velocity range. We describe a criterion of isolation applied to select
our sample, the ``Local Orphan Galaxies'', and discuss their basic optical
and HI properties.
\end{abstract}

\vspace{-0.5cm}
\section{Introduction}
The main goal of our project is to derive a representative sample of
nearby isolated galaxies useful for testing galaxy evolution in low
density regions of the Local universe. At present, there are only two
all-sky samples of isolated galaxies: a list of 197 galaxies in the
Local Volume (= LV, $D <$ 10 Mpc) with negative tidal indexes (Karachentsev
et al., 2004), and a catalog of 3227 2MASS-selected isolated galaxies
(Karachentseva et al., 2010 = 2MIG) compiled in a similar manner as the
catalog of isolated galaxies on the northern sky (Karachentseva, 1973 = KIG).
Both the last catalogs have a typical depth of $\sim$80 Mpc. To fill up
the gap between LV and 2MIG, we create a new sample of ``Local Orphan
Galaxies'' situated in the Local Supercluster within $D <$ 45 Mpc.

\section{Initial data and the isolation criterion}
We tested on isolation $\sim$10500 galaxies with radial velocities in the Local
Group rest frame $V_{LG} < 3500$ km/s situated at galactic latitudes $\mid b\mid > 15\deg$.
First, we cleaned spurious data arrived from automated sky surveys:
SDSS, 2dF, 6dF, DEEP2, etc., then determined $B$-magnitudes and morphological
type if absent. The observed $H,I,R,V,B$- band magnitudes were transformed
into $K_s$ ones to derive the $K$-band luminosity of the galaxies.
 We bound the galaxies in systems based on the following algorithm, which
tooks into account individual properties of galaxies. Thus, two arbitrary
galaxies were considered as a pair if their mutual radial velocity $V_{ik}$
and projected separation $R_{ik}$ satisfy the condition of negative total energy:
 \begin{equation}
  V_{ik}^2 R_{ik}/2GM_{ik} < 1,
\end{equation}
where $M_{ik}$ is the total mass of the pair. We use also the second condition
that the pair components locate inside their ``zero-velocity'' sphere:
\begin{equation}
  \pi H^2 R_{ik}^3/8GM_{ik} < 1,
\end{equation}
where $H_0$ is the Hubble constant. Here we determined masses of galaxies
from their $K$-band luminosity, assuming one and the same ratio
\begin{equation}
  M/L_K = \kappa (M_{\sun}/L_{\sun})
\end{equation}
with a dimentionless parameter $\kappa = 6$. This quantity corresponds
to the mean cosmic ratio of dark-to-luminous matter.

  We identified all pairs satisfying conditions (1)--(3) and then grouped
all pairs with a common component into a single entity. As a result, we
created the catalogs of binary galaxies (Karachentsev \& Makarov, 2008),
triple galaxies (Makarov \& Karachentsev, 2009), and members of groups
(Makarov \& Karachentsev, 2010). This algorithm leaves 46\% of galaxies
as not clusterized, ``field'' ones. Apparently, the higher adopted
quantity $\kappa$, the lower fraction of field galaxies. Increasing the
$\kappa$ in 40 times, we derived a fraction of isolated galaxies to be 10\%.
Finally, we applied to them the Karachentseva's (1973) criterion of
isolation and obtained the sample of 513 Local Orphan Galaxies (LOGs).
 Our subsequent spectral observations of neighbouring galaxies around the
LOGs manifest (Melnyk et al., 2009) that most of them turn out to be
background objects with a median difference of radial velocities +9400 km/s
regarding to LOGs. We conclude that our sample amounts to (85-90)\% true,
spatially well isolated galaxies.

\section{Some properties of the LOG sample}
Distribution of the isolated galaxies on the sky in equatorial coordinates is
shown in Fig.1.

\begin{figure}[!ht]
\begin{center}
\includegraphics[scale=0.5,angle=-90]{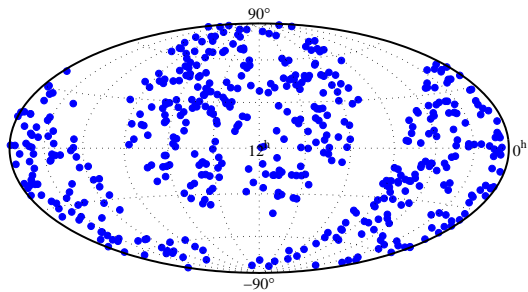}
\end{center}
\caption{Distribution of 513 Local Orphan Galaxies on the sky in equatorial
       coordinates. The Galactic Zone of Avoidance is shaded.}
\end{figure}

The sky distribution looks quite smooth without prominent
over- and under-densities in the regions of known clusters/voids.
Fig.2 presents the distribution of radial velocities of the galaxies
within bins of 250 km/s. The shaded histogram is for galaxies detected in
IRAS.
\begin{figure}[!ht]
\begin{center}
\plottwo{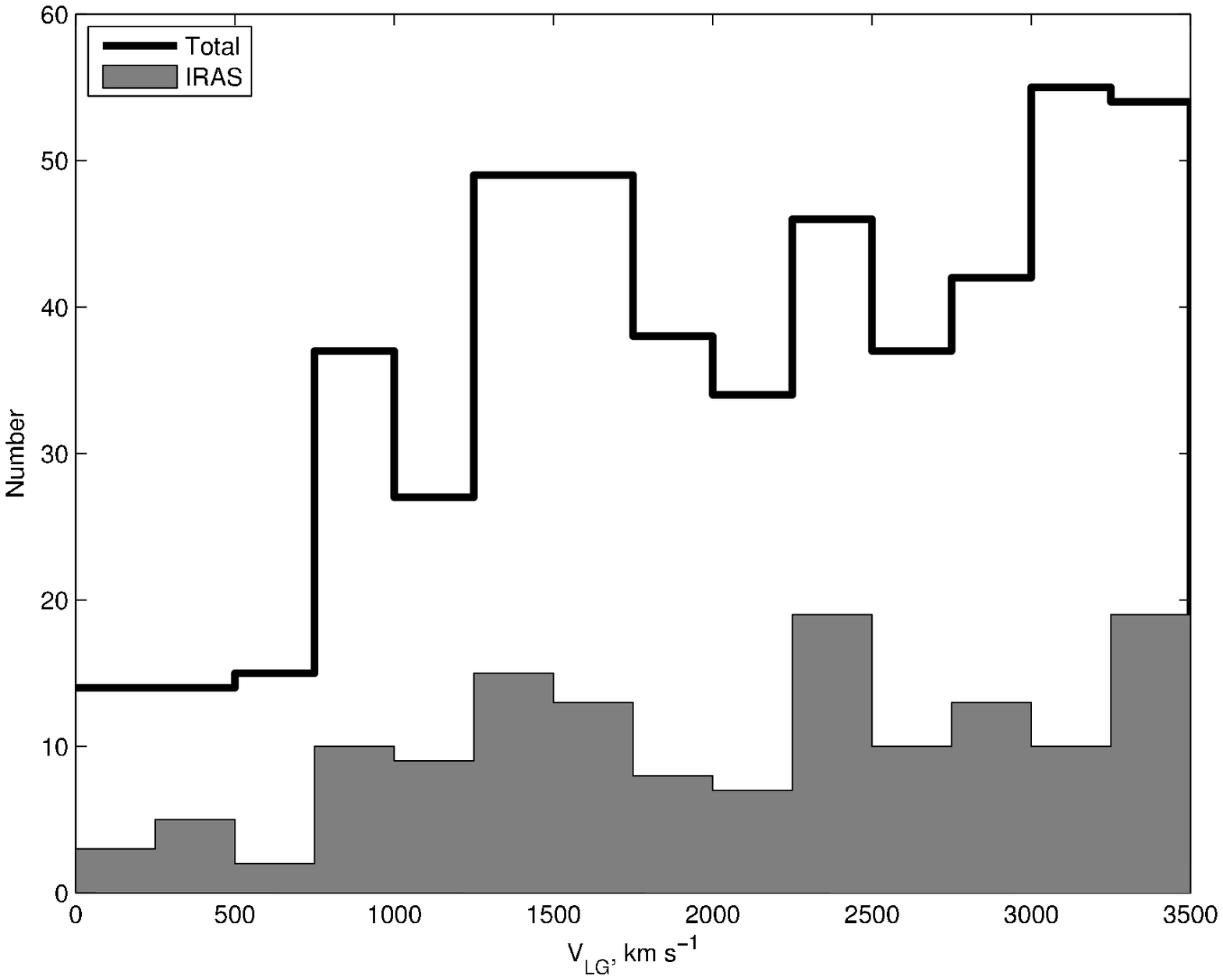}{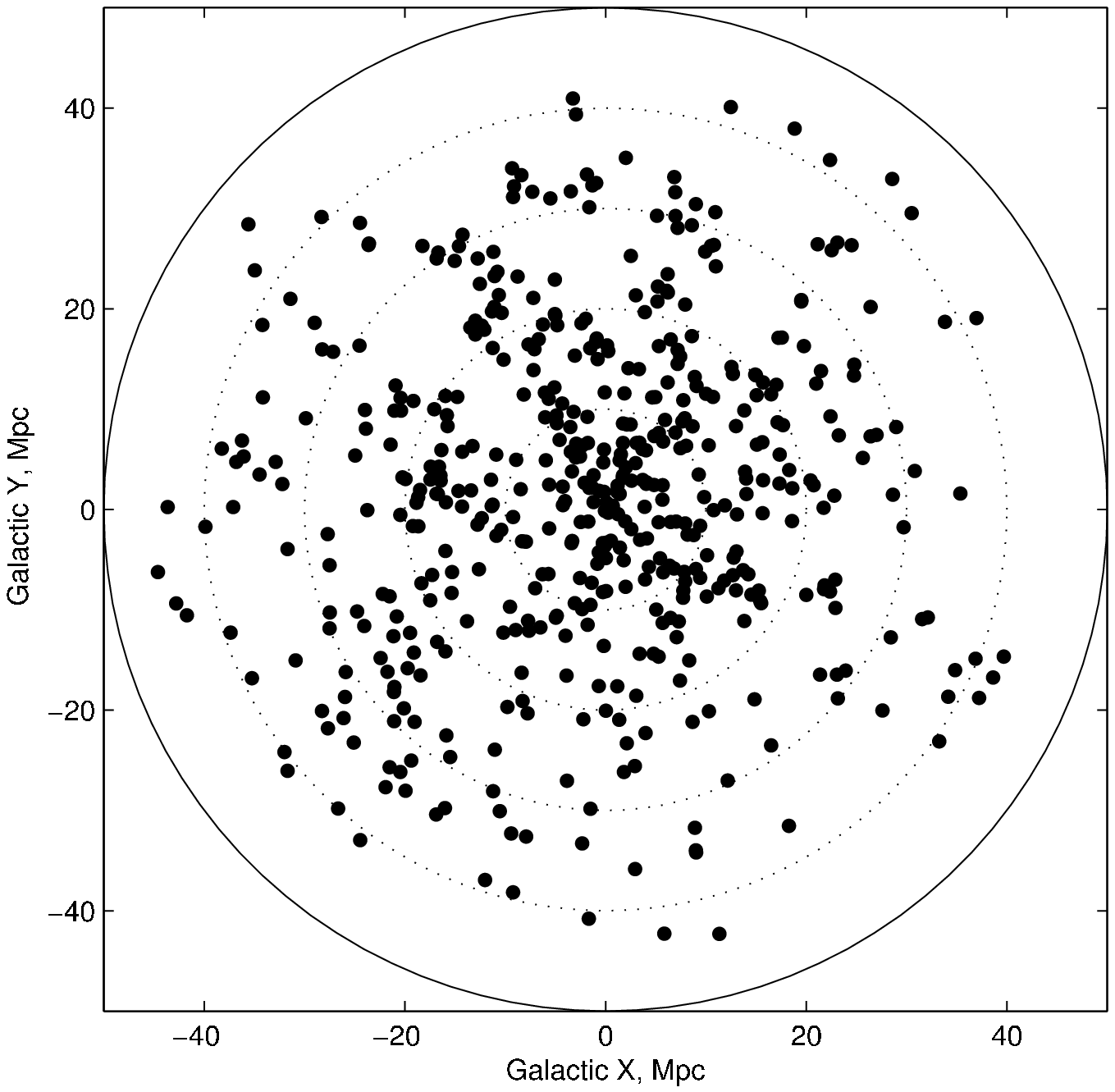}
\end{center}
\caption{Distribution of LOGs on radial velocities with respect to the
       Local Group centroid. Galaxies detected in IRAS are shaded.}
\caption{Spatial distribution of LOGs in projection into Galactic plane}
\end{figure}
Spatial distribution of LOGs in cartesian Galactic coordinates
(Fig.3) exhibits only a moderate clumpiness seen on a scale of 5-10 Mpc.

  The sample of LOGs is dominated by flat, bulgeless galaxies (Fig.4).
About 75\% of the sample are late type objects (T $ > $Sc) with a peak at T = 8 (Sdm).
Notice that the LOG galaxies seen in IRAS have a wider peak at T = 4 -- 6.
Being the distance limited (but not flux-limited) sample, the LOG catalog
is over-represented by dwarf galaxies in comparison with KIG and 2MIG
samples.
\begin{figure}[!ht]
\begin{center}
\includegraphics[scale=0.35]{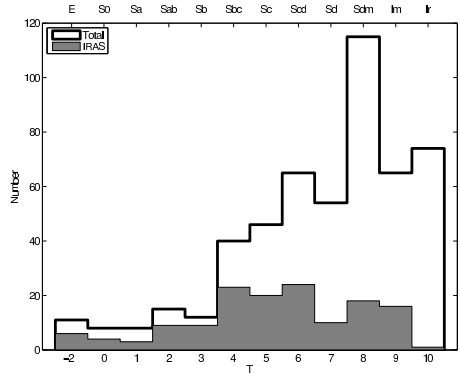}
\end{center}
\caption{Histogram of morphological type distribution for the LOG sample.
       Galaxies detected in IRAS are shaded.}
\end{figure}

  Fig.5 demonstrates distribution of LOGs according to their $K$-magnitudes
and HI- fluxes in the logariphmic scale. The diagonal lines indicate the
total gas-to-stellar mass ratio equal to: 0.01, 0.1, 1, 10 and 100. Most of
the LOGs are gas-rich galaxies with the median $M_{gas}/M_{star} \sim 1$.
\begin{figure}[!ht]
\begin{center}
\includegraphics[scale=0.45]{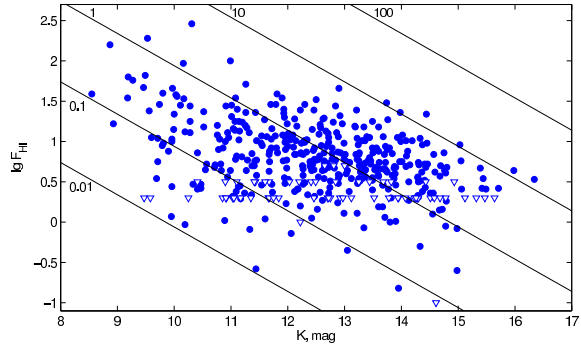}
\end{center}
\caption{Logariphm of HI- flux (in Jy km/s) vs. K- magnitude for the
       LOGs (dots). Galaxies with upper limits of HI- fluxes are shown
       by open triangles. Diagonal lines indicate different levels of
       gas-to-stellar mass ratio.}
\end{figure}

  The E and S0 galaxies, like their nearest representator NGC 404, amounts
to a minor (4\%) fraction of LOGs. They stand out against normal E and S0
galaxies situated in groups and
clusters by a low median luminosity ($M_B = - 17\fm6$)
and the presence of gas and dust.

  We found among LOGs about 20 objects having peculiar structures:
distorted/asymmetric shape or tails that can be interpreted as a result
of recent merging or current interaction with a massive unvisible body
(dark sub-halo). Such well isolated but peculiar objects
deserve closer attention to understand their kinematics and structure.

\acknowledgements This work has been supported by Russ-Ukr
grant 09--02--90414, Ukr-Russ grant F28.2/059, RFBR grant 07-02-00005
and RFBR-DFG grant 06--02--04017.

{}
\end{document}